\documentstyle[pre,aps,epsf,multicol]{revtex}
\begin{document}

\title{ Two-species branching annihilating random walks with one offspring}

\author{Sungchul Kwon and Hyunggyu Park}
\address{Department of Physics, Inha University, Inchon 402-751, Korea}

\date{\today}
\maketitle

\begin{abstract}
We study the effects of hard core (HC) interactions between
different species of particles on two-species branching
annihilating random walks with one offspring(BAW$_2$(1)). The
single-species model belongs to the directed percolation (DP)
universality class. In the BAW$_2$(1) model, a particle creates
one particle of the same species in its neighborhood with the
probability $\sigma (1-p)$ and of the different species with
$(1-\sigma)(1-p)$, where $p$ is the hopping probability. Without
HC interactions, this model always exhibits the DP-type absorbing
transition for all $\sigma$. Even with HC interactions, the
nature of the phase transitions does not change except near
$\sigma=0$, where the HC interaction destabilizes and completely
wipes away the absorbing phase. The model is always active except
at the annihilation fixed point of zero branching rate ($p=1$).
Critical behavior near the annihilation fixed point is
characterized by exponents $\beta = \nu_\perp =1/2$ and $\nu_{||}
= 1$.
\end{abstract}

\pacs{PACS Numbers: 64.60.-i, 05.40.+j, 82.20.Mj, 05.70.Ln}

\begin{multicols}{2}


During last decades, there have been considerable efforts to understand
nonequilibrium absorbing phase transitions from an active phase
into an absorbing phase which consists of absorbing states
\cite{review}. Once the system is trapped into an absorbing state,
it cannot escape from the state forever. Various one dimensional
lattice models exhibiting absorbing transitions have been studied
and most of them turn out to belong to one of two universality
classes, directed percolation (DP) and directed Ising (DI)
universality class. While models with the Ising symmetry between
absorbing states belong to the DI class, models in the DP class
have no symmetry between absorbing states \cite{review,Hwangetc}.

It is generally accepted that the dimensionality and the symmetry
between absorbing states play an important role in determining
the universality classes as in equilibrium critical phenomena. To
find out new universality classes, it is natural to study models
with higher symmetries than the Ising symmetry. To achieve a
higher symmetry such as the Potts symmetry, one may increase the
number of absorbing states. However models with higher symmetries
investigated so far turn out to be always active and only
critical at the annihilation fixed point of zero branching
rate\cite{Hinrichsen}. Recent field theoretical study may provide
a possible explanation for this \cite{CardyTauber}. Although
there is no stable absorbing phase, critical behavior near the
annihilation fixed point are non-trivial and form new
universality classes \cite{CardyTauber,nbaw2}.

In systems with more than two equivalent absorbing states,
there are various kinds of domain walls (or kinks) that cannot cross over
each other to retain the order of
absorbing domains. Multiple occupations of domain walls at a site
is also forbidden. These restrictions may be naturally realized
by introducing hard core (HC) interactions between different domain
walls. These domain wall dynamics can be considered as a
multi-species particle dynamics with HC interactions.

Recently, Cardy and T\"{a}uber introduced an interesting multi-species
model, $N$-species branching annihilating random walks with two
offsprings($N$-BAW(2)), which can be solved exactly for all $N>1$,
using renormalization group techniques in bosonic type formulation
which ignores HC interactions \cite{CardyTauber}.
The $N$-BAW(2) model is a classical stochastic system
consisting of $N$ species of particles, $A_i$ $(i=1,\dots, N)$.
Each particle diffuses on a $d$-dimensional lattice with two competing
dynamic processes: pair annihilation and branching. Pair annihilation is
allowed only between identical particles $(A_i + A_i \rightarrow \emptyset)$.
In the branching process, a particle $A_i$ creates two identical particles
in its neighborhood $(A_i \rightarrow A_i + 2 A_j)$,
with rate $\sigma$ for $i=j$ and rate $\sigma^\prime / (N-1)$ for
$i \neq j$. For $N=1$, this model exhibits a DI-type absorbing transition
at finite branching rate \cite{TT,Jensen,bALR,KwonPark}. For $N>1$,
this model is always active except at the annihilation fixed point of
zero branching rate.

Our recent study on the one-dimensional $N$-BAW(2) model
shows that the HC interaction between different species of particles
drastically changes the universality class in a non-trivial way \cite{nbaw2}.
It contradicts the conventional belief that the HC interaction is
irrelevant to absorbing-type critical phenomena, because the particle
density is so low near an absorbing transition that the probability of
multiple occupations at a site should be too small to be significant.
However, in multi-species models of $N>1$, the asymptotic density
decay near the annihilation fixed point becomes nontrivial
due to the convective displacement generated by the HC interaction.

In this paper, we study two-species branching annihilating
random walks (BAW) with one offspring (BAW$_2$(1)) in one dimension.
The single-species BAW model belongs to the DP universality class \cite{TT,Jensen2}.
So this model is a natural multi-species extension of the DP-type systems.
We numerically study the effect of the HC interaction on the absorbing phase
transition and find that the HC interaction changes the phase
diagram considerably.

The BAW$_2$(1) model is a stochastic system consisting of two species of
particles, $A$ and $B$.
Each particle hopes to a near-neighboring site with probability
$p$ or creates a particle on a near-neighboring site with probability $1-p$.
The created particle can be either of the same species as its parent
with probability $\sigma$ or of the different species with probability
$\sigma^\prime=1-\sigma$. With HC interactions, any dynamics resulting in the multiple
occupations at a site by different species of particles is forbidden.


We perform dynamic Monte Carlo simulations.
With the initial condition of two near-neighboring identical
particles, we measure the survival probability $P(t)$(the
probability that the system is still active at time $t$), the
number of particles $N(t)$ averaged over all runs, and the
mean distance of spreading $R(t)$ averaged over the
surviving runs.

At criticality, these quantities scale algebraically in the
long time limit as $P(t) \sim t^{-\delta}$, $N(t) \sim t^{\eta}$ and
$R(t) \sim t^{1/z}$ \cite{Grass}.
The effective exponents defined as
\begin{equation}
- \delta (t) = \log[P(t)/P(t/b)]/\log b
\end{equation}
and similarly for $\eta(t)$, $1/z(t)$ show the straight line at criticality
and the upward (downward) curvature in the active (absorbing) phase.
With $b=5$, we estimates asymptotic values of dynamic critical exponents
$\delta$, $\eta$, $z$ and critical hopping probability
$p_c$ for various values of $\sigma$.

Fig.~1 shows $\sigma-p$ phase diagram of the BAW$_2$(1) model.
Without HC interactions, this model always exhibits the DP-type
absorbing transition for all $\sigma$. For example, at
$\sigma=0$, we estimate the critical probability $p_c = 0.785(5)$
and dynamic scaling exponents $\delta = 0.155(5)$, $\eta =
0.32(1)$ and $1/z = 0.640(5)$ which agree well with the DP values
\cite{Jensen}.

Even with HC interactions,
the nature of the phase transitions does not change except near $\sigma=0$.
However, near $\sigma=0$, the HC interaction destabilizes and completely
wipes away the absorbing phase. The model becomes always active except
at the annihilation fixed point of zero branching rate ($p_c=1$).
Numerically, we could not pinpoint when the absorbing phase is completely
squeezed out, due to statistical errors. With present simulations,
we estimate the upper bound of $\sigma$ as $\sigma_c\simeq 0.05$.
To identify the scaling behavior near the annihilation fixed point
of the system with the HC interaction, we analyze the finite-size
effects on the steady-state particle density $\rho_s$.

The scaling behavior near criticality is characterized by
exponents $\beta$, $\nu_\perp$ and $\nu_{||}$ defined as
\begin{eqnarray}
\xi   & \sim  \Delta^{-\nu_{\perp}}, \ \ \ \ \ \   \tau  & \sim  \xi^{z}, \nonumber \\
\rho (t)  & \sim  t^{-\alpha},  \ \ \ \ \ \ \rho_{s} & \sim  \Delta^{\beta},
\end{eqnarray}
where $\Delta = p_c - p$, $\xi$ is the correlation length, $\tau$
the characteristic time, $\rho (t)$ the particle density at time
$t$, and $\rho_{s}$ the steady-state particle density.
At the annihilation fixed point, the exponents $\alpha$ and $z$ should follow
from the simple random walk exponents $\alpha = 1/z$ and $z = 2$ \cite{Lee}.

Using the finite-size scaling theory
on the steady-state particle density $\rho_s$ \cite{Aukrust}
\begin{equation}
\rho_s (\Delta,L) = L^{-\beta / \nu_{\perp}} F(\Delta
L^{1/\nu_{\perp}}),
\end{equation}
the value of $\nu_{\perp}$ is determined by collapsing
data of $\rho_s$ with $\beta/\nu_{\perp} = 1$.
We measure $\rho_s$ in the steady state,
averaged over $5\times 10^3 \sim 5\times 10^4$ independent samples
for several values of $\Delta$ ($5 \times 10^{-4} \sim 0.05$) and
lattice size $L$ ($2^5 \sim 2^{9}$).
We find $\nu_\perp = 0.49(4)$ (Fig. 2).
With $\alpha = 1/2$ and $z = 2$, we estimate
\begin{equation}
\beta = 1/2, \ \ \ \  \nu_\perp = 1/2, \ \ \ \  \nu_{||} = 1.
\label{annh}
\end{equation}

The single-species BAW(1) model is one of the simplest model
belonging to the DP class. BAW(1) dynamics involves the
spontaneous annihilation of a particle ($A \rightarrow
\emptyset$) which is a common feature of DP models. A single
particle is spontaneously annihilated by the combination of a
branching (creation) and hopping process such as $ A \rightarrow
A+A \rightarrow \emptyset$.

As $\sigma$ decreases, the pair annihilation by diffusion is hard to
occur, because chance to meet the same species of particles
decreases. It makes the system more active so the critical hopping
probability($p_c$) must be an increasing function of $\sigma^\prime =1-\sigma$.
This is consistent with our numerical phase diagram (Fig.~1).

The effect of the HC interaction on the phase diagram
turns out to be rather tricky.
For large $\sigma$ region, the system tends to form
large $A$- or $B$-dominated domains. The HC interaction
accelerates the pair annihilation process because
the domain boundary will induce the diffusion bias
to the center of each domain. Therefore the system
becomes less active and the absorbing phase expands
with the HC interaction.

However, for small $\sigma$ region, the above story is completely
reversed. The system tends to have a high density of locally
ordered $AB$ pairs.
Without the HC interaction, most of the pair annihilation processes
should occur by diffusions across the domain boundary. So, in this
case, the HC interaction decreases the chance of the pair annihilation
processes and the system becomes more active.
This implies that the critical lines with and without HC interactions
should cross each other like in Fig.~1.

At $\sigma = 0$, a single particle cannot be spontaneously annihilated
by the combination of a single branching process and diffusions. It
needs at least three branching processes, e.g., $A \rightarrow AB
\rightarrow ABA \rightarrow ABAB$, which can be annihilated by
successive diffusions in systems without HC interactions.
So the absorbing phase is still stable and the DP nature is maintained.

However, with HC interactions, the ordered $AB$ pairs can not be
annihilated by diffusions. A single particle has a finite
probability to survive asymptotically and the entire absorbing phase
becomes unstable. The system becomes always active except at the
annihilation fixed point of zero branching rate.

Critical behavior near the annihilation fixed point can be
extracted by balancing the particle density changes due to
branching and pair annihilation processes. Let us consider a
particle $A$ created by a $B$ at time $t$. Annihilation of the
created particle $A$ occurs by encountering an independent $A$
particle through diffusions.  The time scale of this pair
annihilation process is given by ordinary diffusions, i.e.,
$\tau_d\sim d^2$ where the mean distance between particles $d$ is
order of $\rho(t)^{-1}$. Branching process of the parent particle
$B$ increases the particle density with the time scale
$\tau_b\sim (1-p)^{-1}$. By balancing these two time scales, we
expect that the steady-state particle density scales as $\rho_s
\sim (1-p)^\beta$ with $\beta=1/2$. This is consistent with our
numerical finding in Eq.(\ref{annh}).

The above argument should be valid for general $N$-species
BAW(1) models for $N\ge 2$ near the annihilation fixed point.
Preliminary numerical results for $N=3$ confirm our predictions.
The universality class characterized by exponents $\beta =
\nu_\perp = 1/2 $ and $\nu_{||} = 1$ also includes the $N$-species BAW(2)
model for $N\ge 2$ with static branching processes, where two
offsprings are divided by their parent such as $\emptyset B \emptyset
\rightarrow ABA$ \cite{nbaw2}. It is clear that our above argument also
applies to this model.

In summary, we studied the two-species branching annihilating
random walks with one offspring. This model exhibits a
DP-type absorbing phase transition as in the single-species
model. With hard core interactions between different species
of particles, the phase diagram changes considerably.
Especially, the system becomes always active with an annihilation
fixed point for very small branching rate of the same species.
We find that the critical behavior near the annihilation fixed
point is identical to that in the $N$-BAW(2) model with
hard core interactions and static branching.
The $N$-species generalization of the BAW(1) model is currently
under study.


This work was supported by the Korea Research Foundation for the
21st Century and by the research fund through the Brain Korea 21
Project.

\newpage
\begin{figure}
\centerline{\epsfxsize=8cm \epsfbox{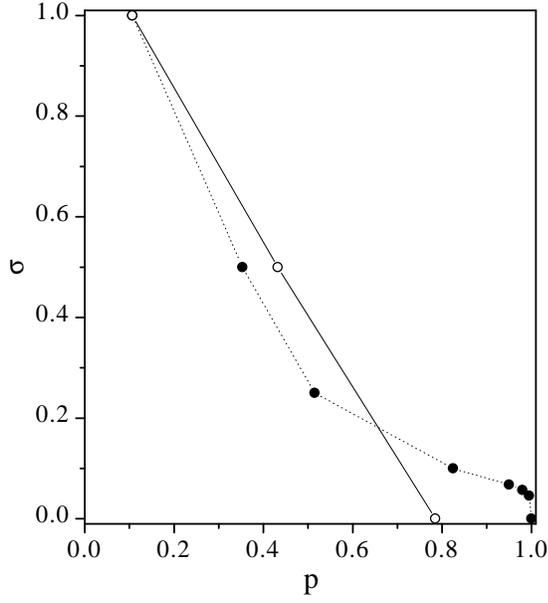}}
\caption{
The $\sigma-p$ phase diagram for the BAW$_2$(1) model.
Open and filled circles correspond to the critical points
without and with the hard core interaction respectively.
Lines between data points are guides to the eyes.
}
\label{fig-1}
\end{figure}

\begin{figure}
\centerline{\epsfxsize=8cm \epsfbox{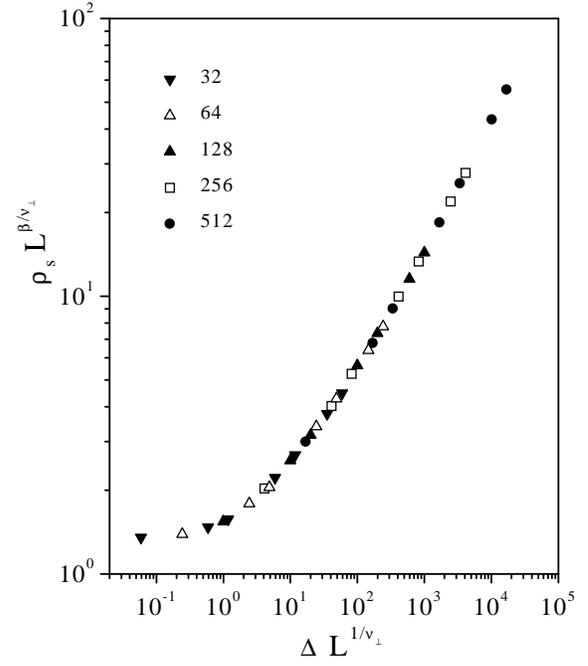}}
\caption{
Data collapse of $\rho_s L^{\beta/\nu_\perp}$
against $\Delta L^{1/\nu_\perp}$ with $\beta/\nu_\perp=1$
for various system size $L=2^5,\ldots,2^{9}$ for
BAW$_2$(1) with the hard core interaction at $\sigma = 0$.
Best collapses are achieved with $\nu_\perp = 0.49(4)$.
}
\label{fig-2}
\end{figure}

\end{multicols}
\end{document}